\documentclass[conference]{IEEEtran}
\usepackage{hyperref}
\usepackage{amsmath,amssymb}
\usepackage{caption,tabularx,booktabs}
\usepackage{graphicx}
\usepackage{epsfig}
\usepackage{epstopdf}
\usepackage{mdwmath}
\usepackage{varwidth}
\usepackage{framed}
\usepackage{float}
\usepackage{placeins}
\usepackage{afterpage}
\usepackage{ragged2e}
\usepackage{amssymb}
\usepackage{pifont}

\usepackage{blindtext}
\usepackage{enumitem}
\usepackage{xcolor}
\usepackage{eqparbox}
\usepackage{fixltx2e}
\usepackage{multirow}
\usepackage{enumitem,lipsum}

\usepackage[english]{babel}
\newcolumntype{P}[1]{>{\centering\arraybackslash}p{#1}}
\usepackage{graphicx}
\usepackage{amsmath, amssymb}
\usepackage{graphicx}
\usepackage{epsfig}
\usepackage{epstopdf}
\usepackage{array}
\usepackage{graphicx}
\usepackage{subcaption}
\usepackage{hyphenat}

\usepackage{booktabs} % For formal tables
\usepackage{graphicx}
\usepackage{caption}
\usepackage{bbding}
\usepackage{pifont}
\usepackage{wasysym}
\usepackage{amssymb}
\usepackage{framed,url,caption,graphicx}

\usepackage{algorithm2e,algorithmic}

\usepackage{soul,xspace,tabularx}
\expandafter\def\expandafter\UrlBreaks\expandafter{\UrlBreaks
  \do\a\do\b\do\c\do\d\do\e\do\f\do\g\do\h\do\i\do\j%
  \do\k\do\l\do\m\do\n\do\o\do\p\do\q\do\r\do\s\do\t%
  \do\u\do\v\do\w\do\x\do\y\do\z\do\A\do\B\do\C\do\D%
  \do\E\do\F\do\G\do\H\do\I\do\J\do\K\do\L\do\M\do\N%
  \do\O\do\P\do\Q\do\R\do\S\do\T\do\U\do\V\do\W\do\X%
  \do\Y\do\Z}

\providecommand{\myparab}[1]{\smallskip\noindent\textbf{#1} }
\newcommand{\cK}{\mathcal{K}}

\newcommand{\eat}[1]{}

\usepackage{physics}
\usepackage[utf8]{inputenc}
\usepackage{algorithm2e}
\SetKwInput{KwData}{Input}
\SetKwInput{KwResult}{Output}
\begin{document}

\title{Performance Optimization in Quantum networks}
\title{Joint Optimization on Entanglement Generation rate 
and Fidelity in Quantum Networks}
\title{EGR Maximization among Organizations with Fidelity threshold}

\title{Path Selection in Quantum Virtual Private Networks}
\title{Flow Optimization in Quantum Networks}
\title{Resource Management in Quantum Virtual Private Networks}
%\title{Flow Optimization in Quantum Virtual Private Networks}
\author{
\IEEEauthorblockN{Shahrooz Pouryousef\IEEEauthorrefmark{1},
Nitish K. Panigrahy\IEEEauthorrefmark{1},
Monimoy Deb Purkayastha\IEEEauthorrefmark{2},
Sabyasachi Mukhopadhyay\IEEEauthorrefmark{2},\\
Gert Grammel\IEEEauthorrefmark{2}, 
Domenico Di Mola\IEEEauthorrefmark{2}, and Don Towsley \IEEEauthorrefmark{1}}
\IEEEauthorblockA{\IEEEauthorrefmark{1}University of Massachusetts Amherst.
% \IEEEauthorrefmark{2}Yale University. 
\IEEEauthorrefmark{2}Juniper Networks. \\
Email: \IEEEauthorrefmark{1}\{shahrooz,nitish,towsley\}@cs.umass.edu, \IEEEauthorrefmark{2} \{monimoyp,sabyas,ggrammel,domenico\}@juniper.net}

}

%\begin{abstract}

%\end{abstract}

% \begin{CCSXML}
% 	<ccs2012>
% 	<concept>
% 	<concept_id>10003033.10003039.10003045.10003046</concept_id>
% 	<concept_desc>Networks~Routing protocols</concept_desc>
% 	<concept_significance>500</concept_significance>
% 	</concept>
% 	</ccs2012>
% \end{CCSXML}

% \begin{CCSXML}
% 	<ccs2012>
% 	<concept>
% 	<concept_id>10002978.10003014.10003015</concept_id>
% 	<concept_desc>Security and privacy~Security protocols</concept_desc>
% 	<concept_significance>500</concept_significance>
% 	</concept>
% 	</ccs2012>
% \end{CCSXML}

% \settopmatter{printacmref=false} % Removes citation information below abstract
% \renewcommand\footnotetextcopyrightpermission[1]{} % removes footnote with conference information in first column
% \pagestyle{plain} % removes running headers

% \keywords{Quantum, routing, entanglement}

\maketitle

\begin{abstract}
In this study, we develop a resource management framework for a \emph{quantum virtual private network} (qVPN), which involves the sharing of an underlying public quantum network by multiple organizations for quantum entanglement distribution. Our approach involves resolving the issue of link entanglement resource allocation in a qVPN by utilizing a centralized optimization framework. We provide insights into the potential of genetic and learning-based algorithms for optimizing qVPNs, and emphasize the significance of path selection and distillation in enabling efficient and reliable quantum communication in multi-organizational settings. Our findings demonstrate that compared to traditional greedy based heuristics, genetic and learning-based algorithms can identify better paths. Furthermore, these algorithms can effectively identify good distillation strategies to mitigate potential noises in gates and quantum channels, while ensuring the necessary quality of service for end users.

%In a public quantum network where a set of flows from different organizations wishes to communicate, path selection is critical to preserve the quality of service requirements of different flows. Distillation at different places of a path may be required to mitigate noises in gates and channels. In this paper, we formulate flow optimization in a quantum network as an optimization problem and use genetic and learning-based algorithms to select a set of paths for multiple organizations. Our results show that genetic and learning-based algorithms can find better paths compared to greedy shortest-hop-based heuristics. In addition, the best distillation strategy can be found by these algorithms to efficiently utilize the resources.

\end{abstract}
\section{Introduction}
The advent of a quantum network will unlock unprecedented possibilities for novel applications, including quantum key distribution \cite{bennett2020quantum, peev2009secoqc,wang2014field,stucki2011long}, distributed quantum computation \cite{cirac1999distributed}, quantum sensing \cite{d2001using}, and clock synchronisation \cite{komar2014quantum}. %Quantum key distribution has further been proven to be successful at short distances in experimental studies \cite{peev2009secoqc,wang2014field,stucki2011long} with a few  commercially available options. 
Crucial to the realization of these applications is the ability to distribute quantum entanglements across long distances. Therefore, the primary responsibility of the first and second generation quantum networks \cite{muralidharan2016optimal} will be to facilitate the delivery of bipartite quantum entanglements, also known as EPR pairs, to distinct pairs of end-users. 

These end-users may belong to various organizations, requiring entanglement distribution service among members belonging to the same organization. Creating individual private quantum networks with dedicated quantum links and switches for each organization may not be scalable due to high infrastructure and operational costs. Thus there exists a critical need for the development of a \emph{quantum vitual private network} (qVPN) architecture that enables multiple organizations to share an underlying public quantum network, analogous to the classical VPNs \cite{Gupta01,Duffield99}. %Entanglement swapping and quantum repeaters are the ingredients of Quantum networks.

\begin{figure}
\centering
\includegraphics[scale=0.41]{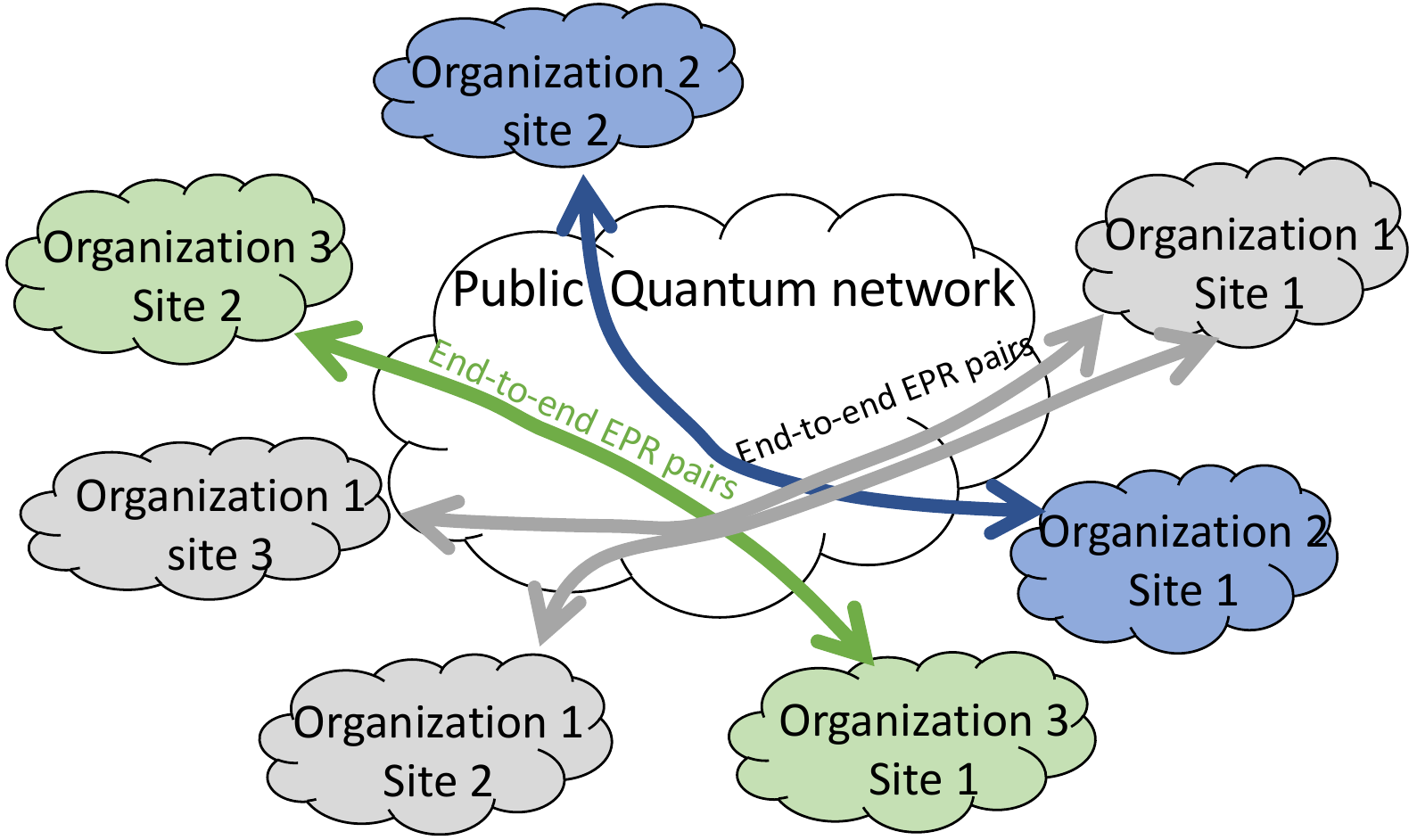}
\caption{A quantum network shared by multiple organizations}
\label{fig:QVPN}
\end{figure}

%\subsection{Flow rate management} 
In this work, we lay out the foundations of a qVPN architecture. Imagine that a group of private organizations need to create intra-organizational EPR pairs between end-users on their sites located in different geographical locations. Figure \ref{fig:QVPN} shows the schematic of such an architecture where an underlying public quantum network is shared among the three organizations. Each organizational site may have an ingress node (i-node) and an egress node (e-node). The responsibility of the public quantum network would be to distribute EPR pairs between the i-nodes and e-nodes of different sites (i-e EPRs), while the organizations themselves would need to create EPR pairs within their respective sites (site EPRs) between end-users and i/e-nodes. Finally, entanglement swap operations would be performed between the site EPRs and i-e EPRs  to create end-to-end EPR pairs (E2E EPRs) between users.

Establishing a  shared quantum network entails resolving a set of fundamental optimization challenges. For example, the E2E EPRs may be consumed by independent applications executed by end-users, enforcing a minimum required value of rate and quality for the generated entanglements. Entanglement generation procedure in a quantum network is inherently lossy and noisy. Thus, to ensure the quality of service requirements, one needs to appropriately allocate the limited underlying quantum network resources to different organizations. Further, organizations may prioritize their members differently when it comes to establishing entanglement. Moreover, the relative importance assigned to each organization in the optimization process may potentially influence entanglement routing, and network resource allocation decisions within the shared public network. One of the goals of this work is to address these challenges to ensure efficient and reliable operation of a qVPN.

In order to optimize the performance of a qVPN, it is also necessary to identify several entanglement generating paths that connect a particular pair of end-users. However, in many cases, there can be an exponential number of simple paths connecting any two nodes in the public  network, making it difficult to consider all possible paths for large networks. In addition, different amounts of entanglement distillation\footnote{Entanglement distillation is a quantum operation that can be performed on low quality EPRs to generate fewer number of high quality EPRs that achieve a target fidelity.} at different places of the network may be required to preserve the fidelity requirements of various organizations and their user pairs. As a result, it becomes crucial to use efficient heuristics to select the most optimal paths and distillation strategies. In this work, we use two heuristic based algorithms: (i) an evolutionary derivative-free Genetic Algorithm (GA), and (ii) a Gradient-based Deep Reinforcement learning (GDR) algorithm for selecting paths and distillation strategies. 

Once a set of paths is selected for every pair of end-users, we tackle an optimization problem aimed at maximizing the total weighted entanglement generation rate (WEGR) for all pairs of end-users across different organizations where different end-user pairs can have different weights, while adhering to link capacity and QoS constraints. Although maximizing WEGR is important, ensuring a fair distribution of rates among end-user pairs prevents any single user or application from monopolizing the available resources. To attain this goal, we introduce constraints on the maximum number of E2E EPR pairs that can be served to each pair of end-users.

Our contributions are summarized below.
% Formulating the above mentioned QON resource allocation problems and experimentally evaluating them are our main contributions in this work. 
% One of the main contributions of this paper is to develop QON resource allocation problem that accounts for optimizing network performance metrics. Our contributions are summarized below.
\begin{itemize}
\item We introduce a quantum virtual private network (qVPN) architecture designed to optimize the distribution of quantum entanglements in a multiorganization setting. To the best of our knowledge, this is the first study to provide a comprehensive design and performance analysis of a qVPN.
\item We present an optimization framework for a qVPN to maximize the aggregate weighted entanglement generation rate, subject to link capacity and QoS constraints.
\item We evaluate the effectiveness of our proposed architecture through extensive numerical simulations. Our results confirm that GA and learning based solution provide $20-30\%$ improvement in W-EGR over the state-of-the-art baselines.
\end{itemize}
The rest of the paper is organized as follows. First, we define the resource allocation problem in qVPN in Section \ref{sec:RA}. In Section \ref{sec:PS}, we explain two path and distillation selection heuristics of our proposed qVPN architecture. An in-depth assessment of the suggested framework's performance is carried out in Section \ref{sec:evaluation}. We list the related work in Section \ref{sec:related} and conclude the paper in Section \ref{sec:concl}.

\section{Resource Allocation in \lowercase{q}VPN}
\label{sec:RA}

\begin{table}
    \setlength{\tabcolsep}{3.5pt}
    \begin{tabularx}{\columnwidth}{c X }
      \toprule
      
      $G(V,E)$& Graph with set of nodes $V$ and set of links $E$
      \\
      \\
      {$c_l$} & {Capacity of link $l\in E$ in EPRs/sec}  \\ 
\\
% $K$ & Set of user pairs\\
        % \\
            $\cK$ & Set of $K$ organizations
    \\
    \\
     $U_k$ $\subset V\times V$ & set of user pairs in organization $k \in \cK$
    \\
    \\
    $w_k$ & Weight of organization $k \in \cK$
    \\
    \\
    $\lambda^k_{u}$ & Weight of user pair $u\in U_k$ in organization $k \in \cK$
    \\
    \\

        {$P^{k}_{u}$}  & {Set of paths connecting  user pair ${u}$ from organization $k$}  \\
        % \\
        % $F_p$ & Basic fidelity of path $p$\\
        %\\
         % $p_u$ & Path for user $u\in \cup_{k\in\cK} U_k$\\
        \\
          {$F^{k}_u$}  & Fidelity threshold of user pair $u$ from organization $k\in \cK$  \\
        \\
        %   {$w^{k}$}  & Weight of user pair $k$  \\
        % \\
        $g_p(F_l, F^{k}_u)$ & Avg. no. of link-level EPR pairs needed for distillation on path $p$ to achieve fidelity threshold $F^{k}_u$\\
        \\
    {$R^{k,u}_{min}$}  & {Minimum rate for user pair $u$ from organization $k$ }   \\
    \\
    {$R^{k,u}_{max}$}  & {Maximum rate for user pair $u$ from organization $k$ }   \\
    \\

    % $\Delta_k$ & Relative weights, set of paths weights
    % \\
    % \\

         {$q$}  & {Swap success probability } 
                   \\ 
\\
\toprule
$x^{k}_{u,p}$ & Entanglement generation rate for user pair $u$ in organization $k\in \cK$ on path $p$
% \\ 
% \\
% $n_{p}^{k}$ & Number of purification rounds on path $p$ for user pair $k$
\\ 
\\
% $g(F_p,n^k_p)$ & the avg number of base EPR pairs used to perform $n^k_p$ number of consecutive successful purification rounds \\
% \\

    %   $f(.)$ & purification scheme
    %   \\
\hline
        
      \bottomrule
    \end{tabularx}
    \caption{\label{table:notaions} Notations used in formulation.}
 %   \vspace{-1cm}
\end{table}

We represent the network as a graph $G(V,E)$ shared by a set of organizations $\cK$ where $V$ is the set of nodes and $E$ the set of links.
Link $l \in E$ generates entanglements (link-level EPRs) with fidelity $F_l$ at rate $c_l$. Each organization $k$ has a set of user pairs $U_k$ that are connected by one or more paths. We associate a weight $w_k$ with organization $k\in \cK$ and each user pair $u$ in organization $k$ has a weight indicated by $\lambda^k_u$. We assume each user pair $u$ has a requirement for maintaining a certain fidelity threshold for the EPR pairs that are provided to them. We represent this fidelity threshold as $F^k_u$. We assume a static fidelity model, meaning that the fidelities of link-level and end-to-end EPRs degrade when entanglement swap, entanglement distillation, measurement, and gate operations are carried out. The notations used in this paper are found in Table \ref{table:notaions}.
%We also assume that the link-level and intermediate EPRs remain coherent when stored in quantum memory.}
%We use $u$ instead of $f$ for flow here as $F$ is used for fidelity. 
%However, one can estimate a lower bound on quantum memory storage time based on the need for classical communication required for swap, distillation and .
\begin{figure*}
\centering
\includegraphics[scale=0.59]{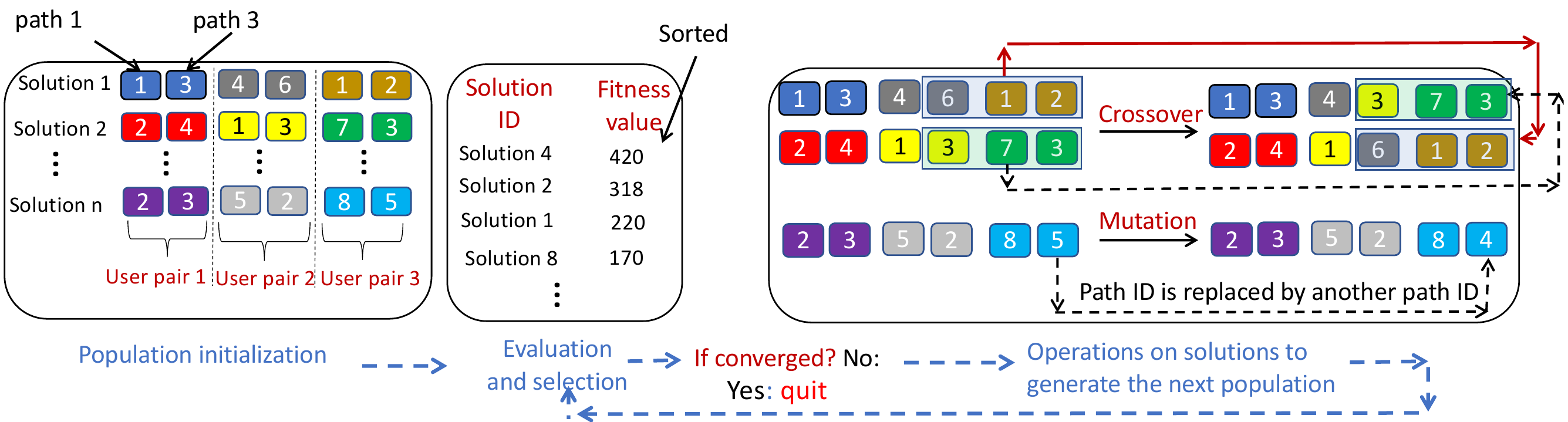}
\caption{A flowchart illustrating the genetic algorithm process for path selection.}
\label{fig:GA}
\end{figure*}

We formulate the resource allocation problem as maximizing the aggregate weighted Entanglement Generation Rate (W-EGR) with respect to satisfying all user pair fidelities $F^k_u$. There are minimum and maximum rate constraints, $R^{k,u}_{min}$ and $R^{k,u}_{max}$ associated with each user pair. Our assumption is that every pair of users $u$, belonging to each organization $k$, has access to a set of one or more paths\footnote{Using Entanglement Swapping, it is possible to establish an E2E EPR  between a pair of users $u$ by utilizing one link-level EPR pair for each link present on the path connecting users $u$.}, denoted as $P^k_u$. 

To satisfy the fidelity threshold requirement of each user, it may be necessary to perform entanglement distillation on each path. Distillation on a path can be performed at two different levels: link-level and end-to-end. In link-level distillation, the fidelities of individual link-level EPRs on a path are enhanced at the expense of some link capacity. Further, the quality of the E2E EPRs can be improved via end-to-end distillation. We employ both link-level and end-to-end distillation in our approach. Let $g_p(F_l,F^k_u)$ represent the average number of link-level EPRs required for distillation on link $l$ having fidelity $F_l$ to achieve a minimum E2E fidelity of $F^{k}_u$, for user $u$ over path $p \in P^k_u$. This function, $g_p(F_l,F^k_u)$, can be computed for various distillation approaches \cite{Deutsch96}, and we will provide further details on this matter in Section $\S$ \ref{sec:distillation_strategies}. We now formulate the following weighted EGR maximization problem for a qVPN.
% \begin{align} 
% \max_{x^k_p}  &\sum_{k\in \cK}\sum_{u\in U_k} \sum_{p\in P^k_{u}} w_k x^k_p \lambda^k_{u} q^{|p|-1}
% \label{problem:flow_optimization}
% % \nonumber
% \\
% \text{s.t.}& 
% \nonumber\\
% & \sum_{\substack{k \in \cK \\
% {u \in U_k}\\
% {p \in P^k_{u}}|l \in p}} x^{k}_p g(F_p,F^k_u)  \leq c_l
% \label{cons:edge_constraint}, \quad l\in E\\
% \quad & x^k_{p}\geq0,  \quad k \in \cK, u \in U_k, p \in P^k_{u}
% \label{cons:variable_constraint}
% \end{align}

\begin{align} 
\max_{x^k_{u,p}}  &\sum_{k\in \cK}\sum_{u\in U_k} \sum_{p\in P^k_{u}} w_k x^k_{u,p} \lambda^k_{u} q^{|p|-1}
\label{problem:flow_optimization}
% \nonumber
\\
\text{s.t.}& 
\nonumber\\
& \sum_{\substack{k \in \cK \\
{u \in U_k}\\
{p \in P^k_{u}}|l \in p}} x^{k}_{u,p} g_p(F_l,F^k_u)  \leq c_l
\label{cons:edge_constraint}, \quad l\in E\\
\quad & \sum_{p\in P^k_u}x^k_{u,p} \geq R^{k,u}_{min} \quad k \in \cK, u \in U_k
\label{cons:min_rate_constraint}
\\
\quad & \sum_{p\in P^k_u}x^k_{u,p} \leq R^{k,u}_{max} \quad k \in \cK, u \in U_k
\label{cons:max_rate_constraint}
\\
\quad & x^k_{u,p}\geq0,  \quad k \in \cK, u \in U_k, p \in P^k_{u}
\label{cons:variable_constraint}
\end{align}

The decision variables in our optimization problem in \eqref{problem:flow_optimization} are $\{x^k_{u,p}\}$ where for each path $p,$ $x^k_{u,p}$ denotes the entanglement generation rate for user pair $u$ from the organization $k \in \cK$ on path $p$ and $q$ is the swap success probability. Here  \eqref{cons:edge_constraint} represents the link capacity constraints, and  \eqref{cons:min_rate_constraint} and \eqref{cons:max_rate_constraint} represent the minimum and maximum rate constraints for each user pair. 

The optimization problem presented in \eqref{problem:flow_optimization} assumes that both the set of paths and the corresponding distillation strategy are already known. We will elaborate on our approach for selecting network paths and the associated distillation strategies in the following section.

\section{Path Selection and Distillation in qVPN}\label{sec:PS}
This section outlines two methods for selecting paths in a qVPN, as well as determining effective distillation strategies for each path to minimize distillation resource expenditure. The selection of a path and corresponding distillation strategy is based on the path's end-to-end fidelity and the required fidelity threshold of the user pair utilizing that path, along with other constraints such as minimum and maximum rate constraints for user pairs. Since it is impractical to compute all possible paths for all user pairs in various organizations and test every possible distillation strategy, we propose the following two distinct schemes. We describe an evolutionary derivative-free approach (via genetic algorithm), as well as a gradient-based approach (via gradient-based deep reinforcement learning) for selecting paths and distillation strategies in a qVPN.

\subsection{Genetic algorithm for path selection}\label{subsec:GA}

The Genetic Algorithm (GA) is a heuristic-based approach for solving optimization problems. %Its process is based on the evolutionary process of a set of individuals observed in nature. 
Initially, GA starts with a population of candidate solutions randomly chosen from the search space of the problem. These candidate solutions undergo successive iterations and a new generation is evolved as a result. Figure \ref{fig:GA} shows how we formulate a candidate solution in our context. 

In our GA approach, a candidate solution is represented as a list with assigned ranges of indices for each user pair. Figure \ref{fig:GA} illustrates this concept, where the first two values of the list correspond to the path IDs for the first user pair. We assume that each user pair can use at most two paths in the network, and the path IDs shown in the figure are randomly selected from a larger set of paths called \textit{candidate paths}. These candidate paths are created for each user pair using heuristic schemes like shortest paths, which we elaborate on in Section $\S$\ref{sec:evaluation}.

Once the initial population is created, each candidate solution is assessed and assigned a fitness value, which indicates its quality. In our approach, the fitness value for each candidate solution is the W-EGR obtained by solving optimization problem \ref{problem:flow_optimization} using the set of paths in that candidate solution as an input. There are three primary operations in a GA: Selection, Crossover and Mutation. To generate new candidate solutions, we apply these operations to each existing candidate solution as follows.
 
% \begin{itemize}
%     \item \myparab{Selection:} determines the probability of using each solution in the current population in the process of generating the next population.

% \item \myparab{Crossover:} indicates with what probability and how two solutions would be combined to generate two new solutions for the next generation.

% \item \myparab{Mutation:} indicates with what probability and how to apply random changes to an individual solution to create a new solution.

% \end{itemize}

In Figure \ref{fig:GA}, the GA operates on a set of $n$ candidate solutions. Candidate solutions $1$ and $2$ are first selected for the crossover operation. As shown, the first three cells of candidate solution $1$ are moved directly to the first (top) new candidate solution in the next generation, and the second half of candidate solution $2$ is moved to the second part of the first new candidate solution. For the mutation operation, the last cell in candidate solution $n$ is selected. Subsequently, the GA evaluates the fitness values of the new population.

\subsubsection{Exploration vs. exploitation in GA}
GA applies mutation and crossover operations to create new candidate solutions. In the \emph{dynamic version} of GA, the algorithm adapts its policy for these operations over time to balance exploration of the search space with exploitation of what has been discovered so far. More specifically, the dynamic version of GA tends to explore the search space more in the early generations and reduces exploration as time passes. This is controlled by adjusting the probability of mutation and crossover and the selection operation. For example, the dynamic GA may give most of the candidate solutions in the early generations a chance to generate the next population, while using higher values for the probability of mutation and crossover. In contrast, the static version of GA always uses the same strategy for selection, such as using a fixed subset of candidate solutions to generate the next population. We assess the effectiveness of both static and dynamic approaches via numerical simulations in Section \ref{sec:evaluation}.
% The other and last difference between our two versions of GA is that the dynamic version of GA would apply mutation and crossover at higher points of a chromosome at early generations and reduce this over time. The static version will always use a fixed value for the probability of doing these operations.

\subsection{Gradient-based reinforcement learning for path selection}\label{subsec:RL}

We now employ a neural network as a policy function, which can be adjusted to map the given set of user pairs to a probability distribution representing the likelihood of selecting each path for every user pair. Unlike genetic algorithms (GA), this method is based on derivatives and we formulate the learning aspect of this approach as a reinforcement learning problem.

\begin{figure}
\centering
\includegraphics[scale=0.45]{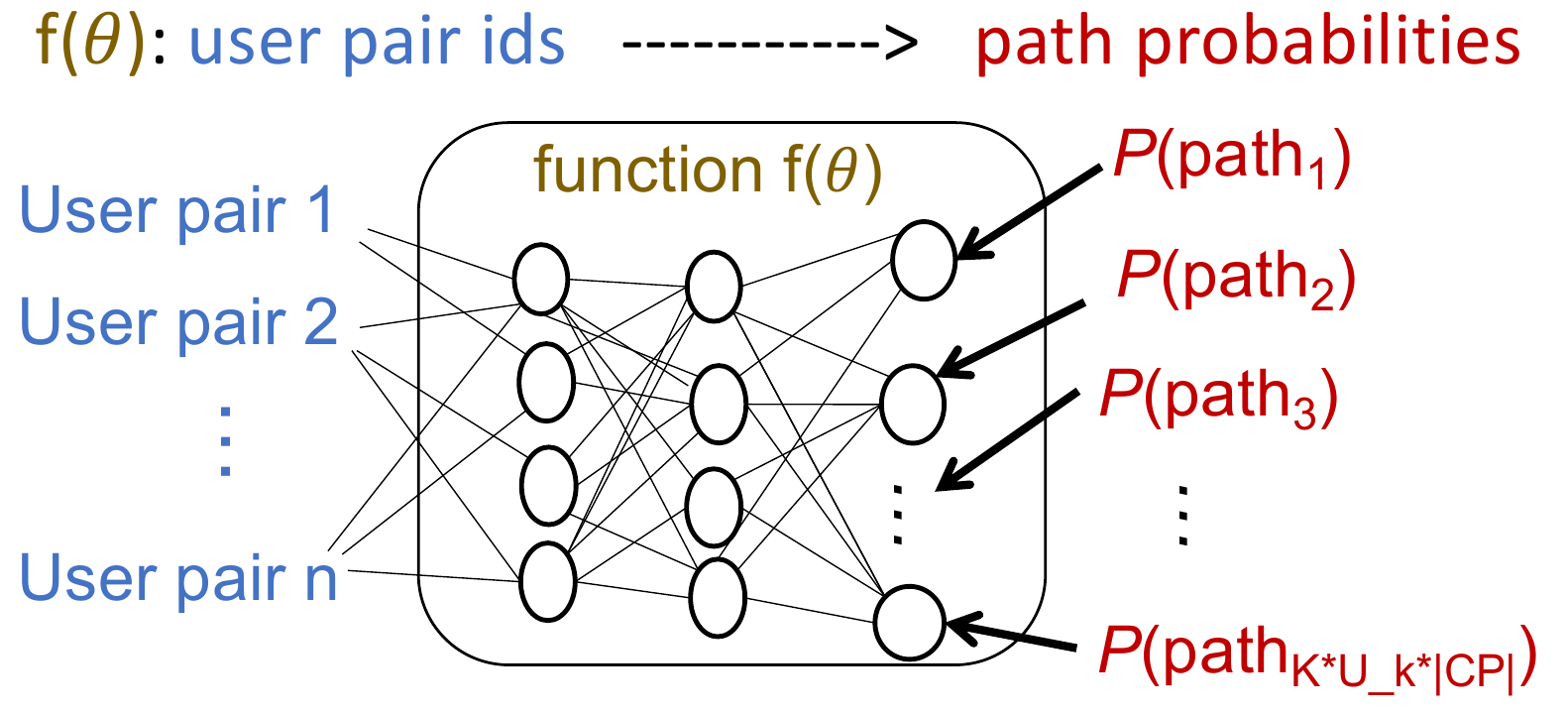}
\vspace{-0.06in}
\caption{Representing the function of path selection as a neural network.}
\label{fig:gb_for_path_selection}
\end{figure}

Figure \ref{fig:gb_for_path_selection} depicts how a neural network is used to learn a set of paths for our optimization problem. Path selection is obtained by training the neural network, where the input is a list of user pairs and the output is the probability of selecting each path in the network. If there are $K$ organizations and $U_k$ user pairs in each organization, and $|CP|$ candidate paths for each user pair, the output size of the neural network is $KU_k|CP|$. From this output, we select $R$ paths, where $R$ is equal to $KU_k|P|$, and $|P|$ is the maximum number of paths allowed to be used by each user pair ($|P|\leq |CP|$). We select the paths with the largest probabilities as inputs to problem \eqref{problem:flow_optimization} and calculate the aggregate weighted EGR. We then use the computed W-EGR to update the weights of our neural network and improve the neural network model. We repeat this process to derive a function policy that maps the set of user pairs to an output indicating the paths that maximize the aggregate W-EGR. Further details about the training algorithm are provided in Appendix \ref{sec:appdx}, Algorithm $\S$\ref{alg:training_algorithm}.

\subsection{Distillation strategies} 
\label{sec:distillation_strategies}

A distillation strategy for a path involves determining the appropriate amount of link-level and end-to-end distillation necessary to attain the desired E2E fidelity level. For example: consider a path $p$ used to create E2E EPR pair between user pair $u$ for organization $k.$ For a path with a length of $N$ and identical link fidelities set to $F_L$, the resulting fidelity of the E2E EPR pair between $u$ after undergoing $N$ nested swap operations can be expressed as
%\footnote{We approximate the state of each LLE $\rho$ as a Werner state \cite{werner1989quantum} with the same fidelity, i.e. $F_L = 0.8$. }:
\begin{equation}\label{eq_nested_swaps_equal}
     S(F_L) = \frac{1}{4} + \frac{3}{4}{(\frac{P_2(4\eta^2-1)}{3})}^{N-1} {(\frac{4F_{\text{L}}-1}{3})}^{N}
\end{equation}
 
\noindent where  $P_2$ is the two-qubit gate fidelity and $\eta$ is the measurement fidelity of the swapping operation. Let $F_{out} = \tilde{S}(F_{in})$ represent the relationship between input and output fidelities of a distillation algorithm. A distillation strategy for path $p$ can then be defined as a fidelity threshold for link-level EPRs (say $F_L^{th} > F_L$) such that $\tilde{S}(S(F_L^{th})) \ge F_u^k.$ The format of the candidate solution (or output) in GA (or Gradient-based RL) in Section \ref{subsec:GA} (or \ref{subsec:RL}) can be modified to include both $p$ and $F_L^{th}$ for joint path and distillation strategy selection.

%The format of the candidate solution in  to for path and distillation strategy selection.
%determined by a fidelity threshold that needs to be met for the link-level EPR pairs of that path. 

% By this, we can have an infinite number of distillation strategies for a path. Each distillation strategy is a combination of the edge-level and end-level distillations. For each distillation strategy, first, the link-level EPR pairs are distilled until the fidelity threshold of the distillation strategy for the link-level EPR pairs is satisfied and then we do distillation at the end-to-end level if the E2E path fidelity after performing edge-level distillation does not satisfy the fidelity threshold of the user pair using that path. The amount of distillation to perform at each edge to reach a certain edge-level fidelity is determined by the distillation protocol we use. 

%\myparab{Encoding distillation strategies in GA solution:} One way to encode using different distillation strategies in a GA solution is to have different versions of a path and let the GA select among them. In other words, for each path $p$ and for each distillation strategy, we consider a unique id that whenever that ID is selected by the GA, we perform the amount of edge-level and end-level on that path regarding the distillation strategy of that ID. 

\section{Evaluation}
\label{sec:evaluation}
In this section, we conduct a comprehensive set of experiments aimed at assessing various approaches for selecting path and distillation strategies in a qVPN. We have chosen three different baselines  based on the shortest paths between user pairs. These schemes use the shortest paths in the network for each user pair when the link weights in the network are set to one (for Shortest hop-based scheme), to $1/\text{EGR}$ (for Shortest EGR-based scheme) and $1/\text{EGR}^2$ (for Shortest EGR-Square-based scheme) with EGR being the link capacity. For each user pair in each organization, we compute the set of candidate paths for GA and Gradient-based scheme using the $k$-shortest path algorithm proposed in \cite{yen1971finding} with $k = 5.$ Our goal is to address the following research questions: (1) How do the performance of GA and the \textit{Gradient-based} scheme compare to that of the alternative greedy baseline methods? (2) How do search space and problem size affect the convergence of GA and the gradient-based scheme? (3) Can GA effectively identify suitable distillation strategies for various paths and different fidelity thresholds  in order to mitigate noisy gates and meet user requirements? (4) How can we ensure fair resource allocation in a qVPN?

% \begin{table}
%     \centering
%     \setlength{\tabcolsep}{4.6pt}
%     \small
%     \begin{tabularx}{\columnwidth}{L L L L L}
%         \toprule
%         Param & $\cK$  & $w_k (\FORALL{k\in \cK})$& $\lambda^k_u (\FORALL{k\in \cK, u\in U_k})$  & $F^{k}_u (\FORALL{k\in \cK, u\in U_k})$ \\ 
%         \midrule
%         Value & $3$ &Unif$[0.1,1.0]$&Unif$[0.1,1.0]$&Unif$[0.75,0.94]$\\
%         \bottomrule
%     \end{tabularx}
%     \vspace{-0.05in}
%     \caption{\label{tab:parameters_value} Parameters used in our experiments.}
% \end{table}

\myparab{Network topology:} In our evaluation, we use the Dutch SURFnet network, taken from the Internet topology zoo \cite{zoo}. We obtain the geographical location of the nodes in this network from the zoo topology data set \cite{zoo} and compute the length of each link accordingly. %We assume there is a direct link between two nodes of an edge on this network.

\myparab{Computing link capacities:}
We assume link-level EPR pairs are generated using a single-photon scheme \cite{humphreys2018deterministic}. In addition, we assume one can control the rate and the fidelity of the generated EPR pairs on each link with a tunable parameter $\alpha$ \cite{humphreys2018deterministic}. In other words, we assume that each LLE is modelled by a state of the form $\rho= (1 - \alpha)|\Psi\rangle \bra{\Psi^+} +\alpha |\uparrow \uparrow\rangle \bra{\uparrow\uparrow}$
with probability $p =  2\eta_l \alpha$ where $\eta_l$ is the transmissivity for link $l$ with length $d_l$.  $\eta_l$ is computed as $\eta_l=10^{-0.1\beta d_l}$ with $\beta = 0.2$ dB/km and is the fiber attenuation coefficient.  Then, we compute the rate as $\frac{p}{T}$ where $T$ is the repetition time. We set $T$ to one $\mu s$ and $\alpha$ to $0.2$ in our experiments, resulting in the link fidelities being configured to $0.8$.

\myparab{Engineering the SURFnet topology:} Some of the longer links in the SURFnet topology produce significantly lower EPR rates compared to other shorter links. For this reason, we engineer this topology by adding a repeater every 10km along links exceeding 20km in length. However, we do not allow these newly added repeater nodes to be used as end points for users. Additionally, we assume a multiplexing factor of three for each link, effectively tripling its capacity.

% \eat{We set the rate of entanglement generation on each edge in all networks to a value randomly chosen between 200 and 1400.  The fidelity of each edge in the network is randomly chosen from the range [0.96,0.99].
% We set the number of time intervals to 10 ($|T|=10$) and the duration of each time interval ($\Delta$) is 20 seconds.}

\myparab{Purification protocol:} In our experiments, we use the recurrence-based purification scheme \cite{dur1999quantum}. In particular, we use the DEJMPS protocol \cite{Deutsch96} for purification. The values of the function $g(.)$ for different initial fidelity and target fidelity thresholds can be determined from the results presented in \cite{Deutsch96}. In our experiments, we assume $P_2 = 1$ and $\eta = 0.99$.
% \begin{equation}\label{eq_nested_swaps_unequal}
% \small
% \begin{split}
%     F = \frac{1}{4} + \frac{3}{4}&{\biggl(\frac{P_2(4\eta^2-1)}{3}\biggl)}^{N-1}\times\biggl(\frac{4F_1-1}{3}\biggl)\\
%     &\times\biggl(\frac{4F_2-1}{3}\biggl)...\biggl(\frac{4F_N}{3}\biggl),
% \end{split}

\begin{table}[t]
\footnotesize
\begin{small}
    \centering
    \small
    \footnotesize
    \begin{tabularx}{\columnwidth}{X c c c c }
      \toprule
      {Param} &   {$\cK$} & {$w_k$}&$\lambda^k_u$ &$F^k_u$ \\
    \midrule
      {Value} & {$3$}  &  {Unif$[0.1,1.0]$} &Unif$[0.3,0.7]$& {Unif$[0.75,0.90]$} 
    \\
      \bottomrule
    \end{tabularx}
    \vspace{-0.02in}
    \caption{\label{tab:parameters_value} Parameters used in our experiments.}
    \vspace{-0.1in}
    \end{small}
\end{table}

\begin{figure*}
\centering
\begin{subfigure}{.32\textwidth}
    \centering
    \includegraphics[width=1.0\linewidth]{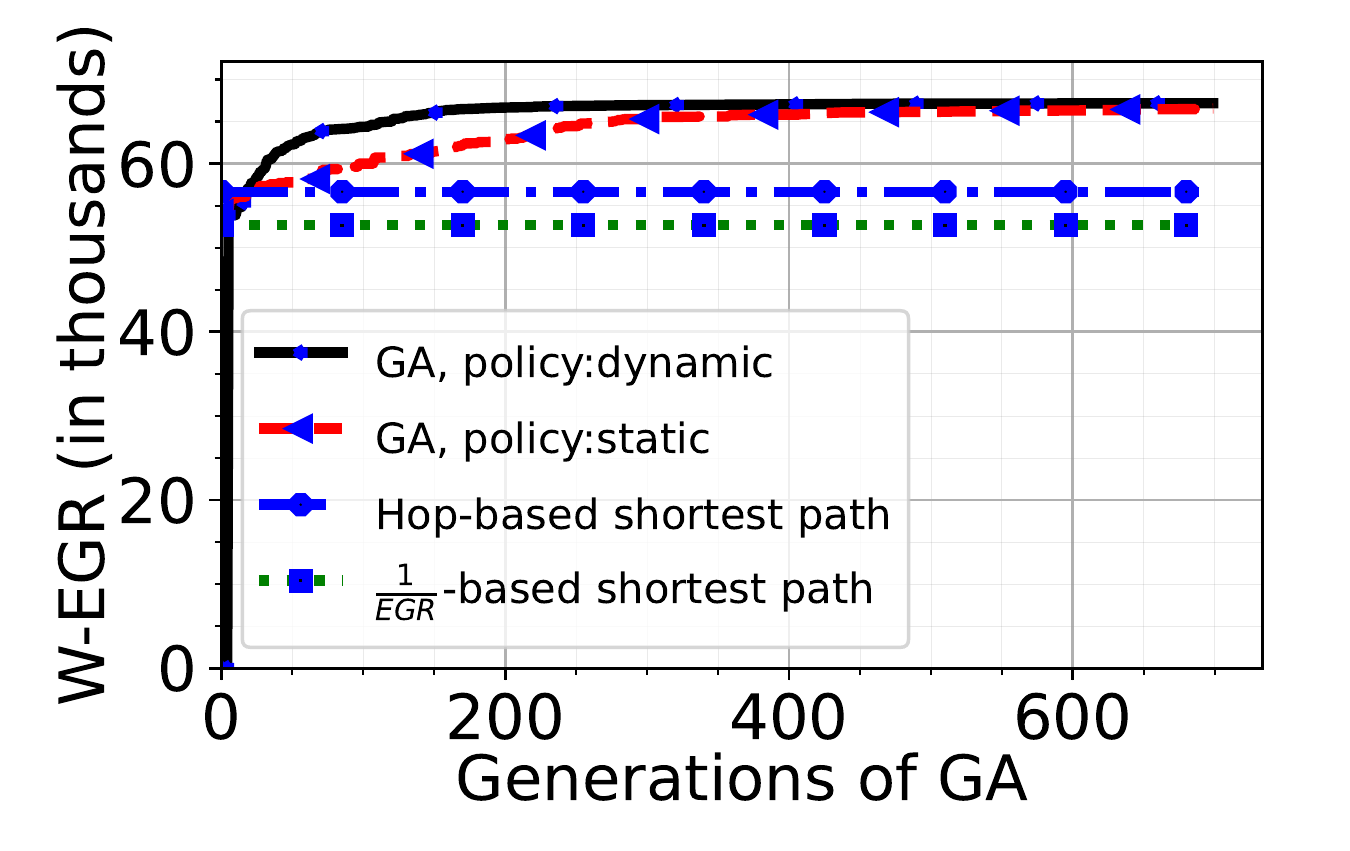}  
     % \vspace{-0.06in}
    \caption{GA vs. baseline algorithms}
    \label{fig:QVPN_GA_and_greedy_schemesv4}
\end{subfigure}
\begin{subfigure}{.32\textwidth}
    \centering
    \includegraphics[width=1.0\linewidth]{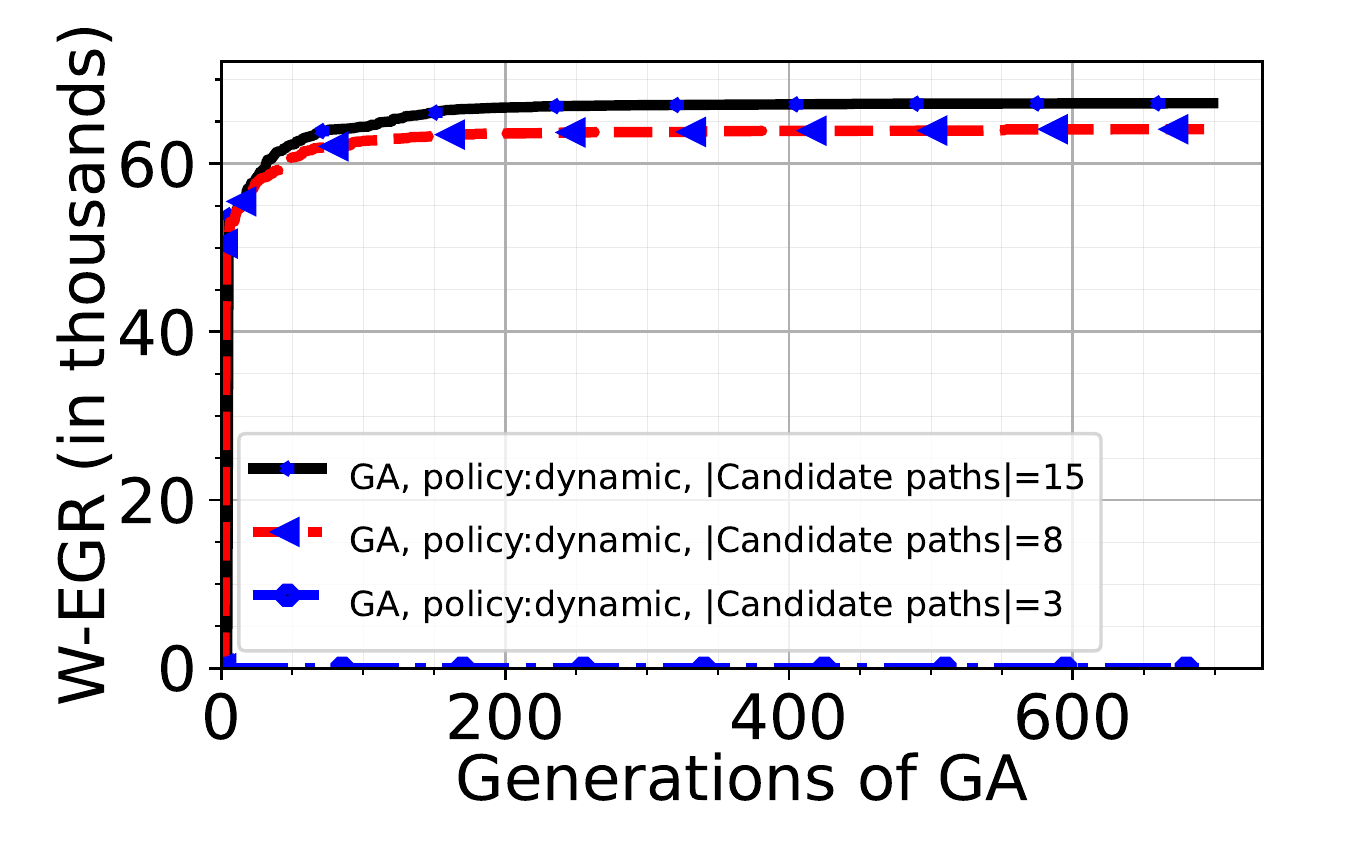} 
    % \vspace{-0.06in}
    \caption{Different number of can. paths}
    \label{fig:QVPN_GA_affect_of_search_space_candidate_paths}
\end{subfigure}
\begin{subfigure}{.32\textwidth}
    \centering
    \includegraphics[width=1.0\linewidth]{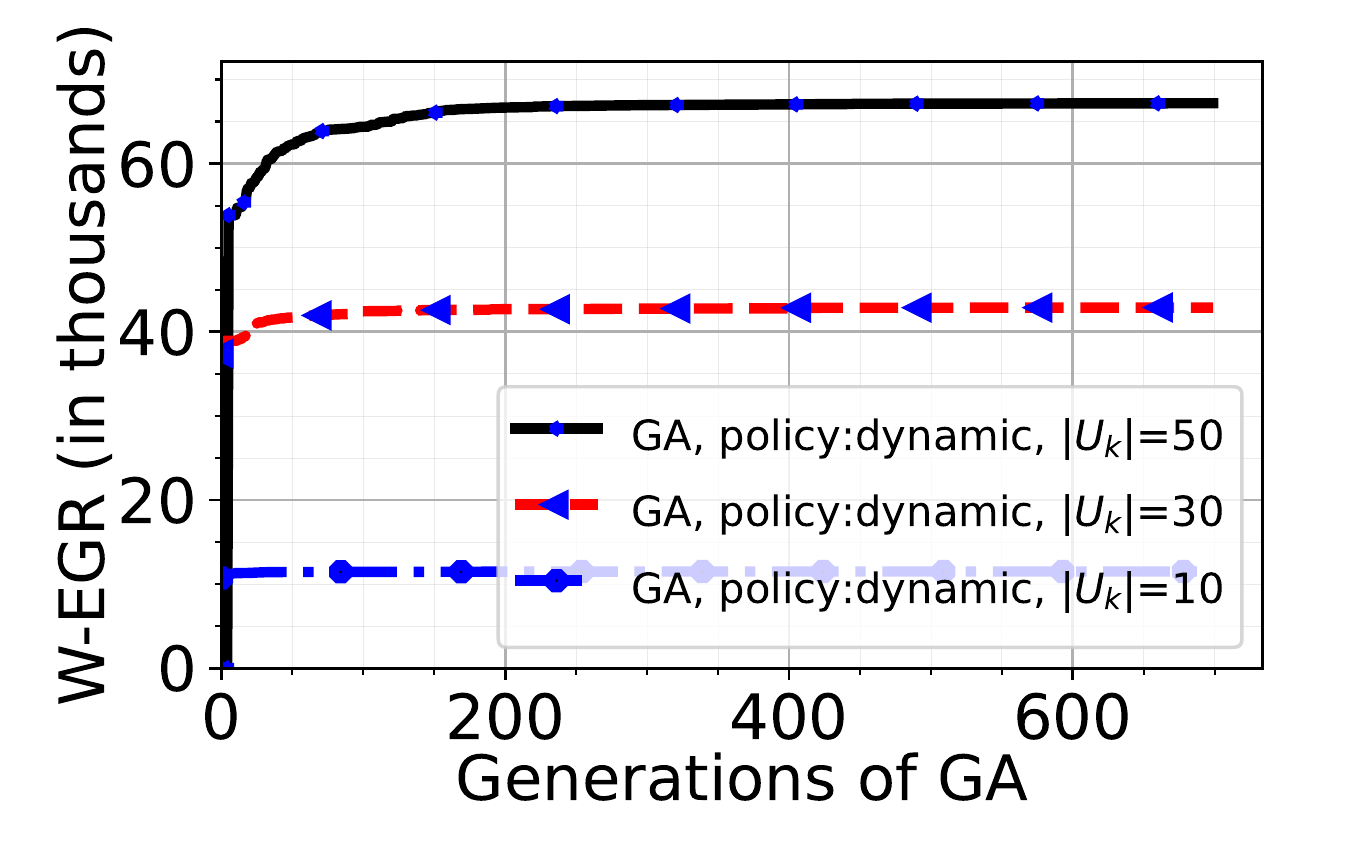}
    % \vspace{-0.06in}
    \caption{Different number of user pairs}
    \label{fig:QVPN_GA_affect_of_search_space_flows}
\end{subfigure}
%\vspace{-0.15in}
\caption{Evolution of GA over 1000 generations in different setups. GA over-performs the baseline algorithms (\ref{fig:QVPN_GA_and_greedy_schemesv4}), explores the available paths as candidate paths successfully (\ref{fig:QVPN_GA_affect_of_search_space_candidate_paths}), and is able to handle different problem sizes (\ref{fig:QVPN_GA_affect_of_search_space_flows}).  \label{fig:evolution_of_egr_in_genetic_algorithm_with_different_dis_alpha_gate_noise}}
\end{figure*}

\myparab{Workloads} We assume there are three organizations in a qVPN and each organization has $50$ user pairs unless noted otherwise. We randomly select the user pairs in the network among the nodes of the original SURFnet topology and set their fidelity threshold uniformly from the interval of $[0.75,0.90]$. We select nodes whose shortest hop-based path length is no more  than $7$. Table \ref{tab:parameters_value} shows the value of different parameters used in our experiments.  In Figures \ref{fig:QVPN_GA_and_greedy_schemesv4}, \ref{fig:QVPN_GA_affect_of_search_space_candidate_paths},  \ref{fig:QVPN_GA_affect_of_search_space_flows}, and Figure \ref{fig:QVPN_affect_of_using_multiple_paths_per_flow} we assume $R^{k,u}_{min} = 10$ for all user pairs in each organization  and $R^{k,u}_{max}=1,000$ unless otherwise specified. We consider four distinct workloads and plot the results of GA as the average of two runs for each of them. Each workload indicates 50 user pairs for three different organizations with varying  fidelity thresholds and user pair weights. 

%Two user pairs from two different organizations can be running on the same end nodes. 

In all our experiments, we  solve  optimization problem \ref{problem:flow_optimization} using the IBM CPLEX solver \cite{cplex}.

% \myparab{Greedy-based baseline algorithms:} We have chosen three different baselines  based on the shortest paths between user pairs. These schemes use the shortest paths in the network for each user pair when the link weights in the network are set to one (for Shortest hop-based scheme), to $1/\text{EGR}$ (for Shortest EGR-based scheme) and $1/\text{EGR}^2$ (for Shortest EGR-Square-based scheme) with EGR being the link capacity. For each user pair in each organization, we compute the set of candidate paths for GA and Gradient-based scheme using the $k$-shortest path algorithm proposed in \cite{yen1971finding} with $k = 5.$ %is $5$ following these heuristic schemes. 

\myparab{Distillation strategies:} We consider $16$ distillation strategies for each path in our experiment. These strategies are constructed by selecting the link-level fidelity thresholds from a fixed set of fidelities ranging between $0.8$ and $0.998.$ We use a link-level fidelity threshold of $0.992$ for the greedy-based heuristic schemes.
%$\{0.8,0.85,0.88,0.89,0.9,0.92,0.94,0.96,0.97,0.98,0.988,0.990,\\0.992,0.994,0.996,0.998\}$. 

\myparab{Implementation:} We develop our GA algorithm in Python and the gradient-based scheme in Python and TensorFlow \cite{abadi2016tensorflow} framework. For the gradient-based scheme, our policy neural network has three layers. The first layer is a convolutional layer with $128$ filters. The kernel size is $3 × 3$ and the stride is set to 1. The second layer is a fully-connected layer with 128 neurons. We use Leaky ReLU for the activation function for the first two layers. The final layer is a fully connected linear layer (without activation function) with $N*(N - 1)$ neurons corresponding to all possible sets of user pairs in the network. The softmax function is applied to the output of the final layer to generate the probabilities for all available actions. The learning rate is initially configured to $0.0001$ and decays every $500$ iterations with a base of $0.96$ until it reaches the minimum value of 0.0001.

\subsection{GA vs. baseline algorithms:}

We first compare two variants of GA against the three baselines to determine their effectiveness. The two versions of GA are distinguished by their exploration and exploitation policies. Figure \ref{fig:QVPN_GA_and_greedy_schemesv4} shows the performance of dynamic and static versions of GA compared to the three baselines. We constrain each user pair in each organization to use a maximum of 3 paths.
%each user pair in each organization is allowed to use at most 3 paths and the size of the candidate paths for each user pair is potentially 15 multiplied by the number of distillation strategies that we consider for each path. 
We plot the aggregate weighted entanglement generation rate (W-EGR) as a function of the number of generations of the GA. The plot demonstrates that even if GA commences with a population of solutions inferior to those generated by the baselines, it eventually converges to a better solution. We observe that the dynamic version of GA converges faster as compared to the static version since the latter does not adequately explore the search space at the beginning. We will employ dynamic version of GA in later experiments. 

%We tried different distillation strategies for the three greedy-based schemes and found that these schemes with some distillation strategies may not even be able to satisfy the constraints of minimum rate for the user pairs.   

% and \ref{fig:QVPN_affect_of_using_multiple_paths_per_flow}

\subsection{GA vs. Search Space}
We now investigate the impact of search space size and problem size on the convergence rate of GA and quality of the solution. In particular, we explore how the size of the search space for paths and the number of user pairs in each organization affects convergence of the GA. We hypothesize that a smaller search space will lead to faster convergence, while a larger search space will lead to slower convergence but a better solution. This pattern is indeed what we observe in Figures \ref{fig:QVPN_GA_affect_of_search_space_candidate_paths}, \ref{fig:QVPN_GA_affect_of_search_space_flows}. 
As the number of candidate paths increases (Figure \ref{fig:QVPN_GA_affect_of_search_space_candidate_paths}), GA is able to choose paths from a larger set, resulting in a better solution. However, it takes more generations to converge. Similarly, when the number of user pairs in each organization increases (Figure \ref{fig:QVPN_GA_affect_of_search_space_flows}) GA requires more generations to find better paths for each user pair. % due to the higher number of genes in each individual.

\subsection{Benefit of multiple paths in a qVPN}

\begin{figure}
\centering
\includegraphics[scale=0.53]{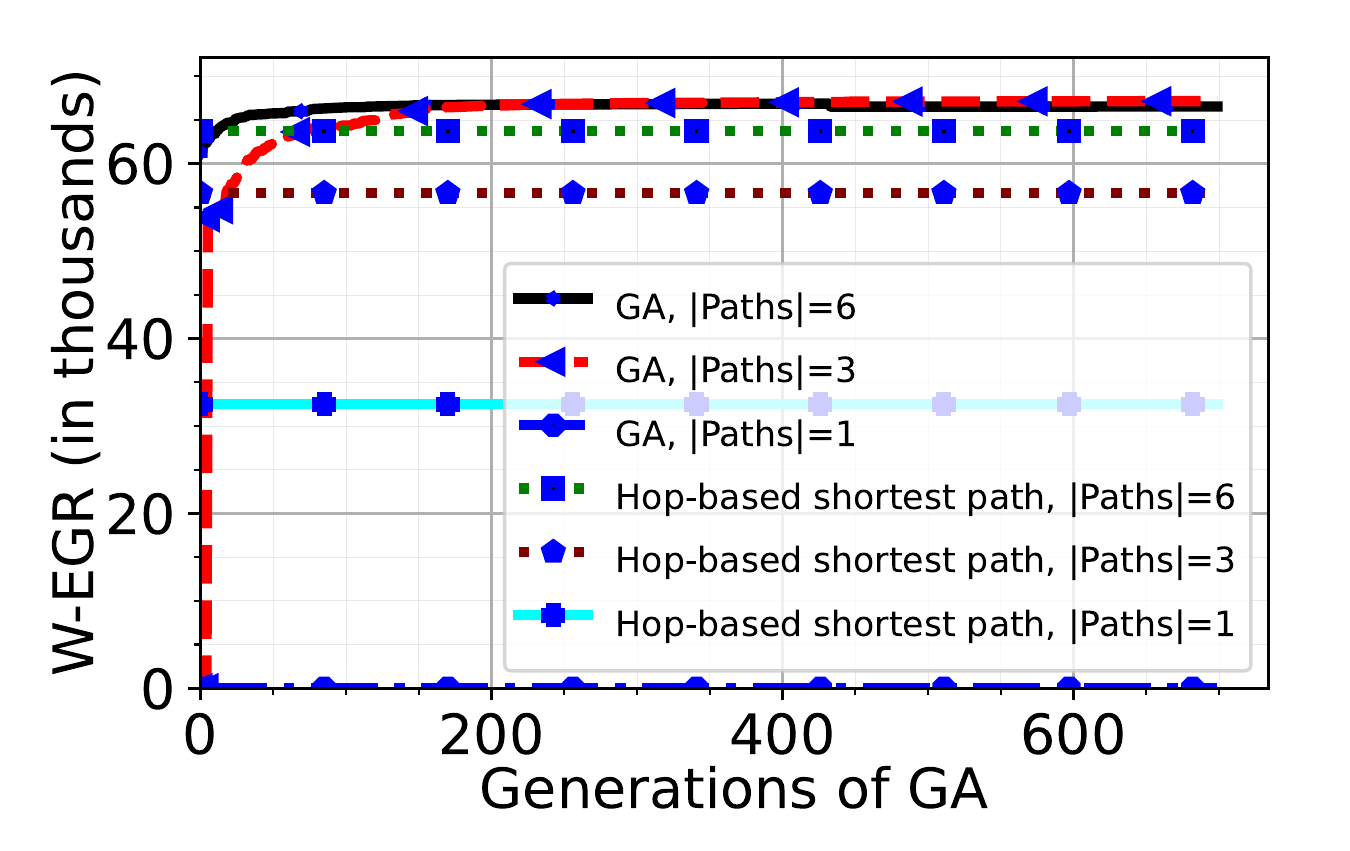}
%\vspace{-0.15in}
\caption{Effect of using multiple paths in a qVPN (zero W-EGR indicates the infeasibility of solving the optimization problem).}
\label{fig:QVPN_affect_of_using_multiple_paths_per_flow}
\end{figure}

Now we investigate the benefits of using multiple paths in a qVPN.  Figure \ref{fig:QVPN_affect_of_using_multiple_paths_per_flow} illustrates that utilizing multiple paths can offer higher rates than a single path. Moreover, using multiple paths can aid GA and the baselines in finding feasible solutions to the optimization problem. Specifically the figure shows that, when the number of paths per user pair is one, W-EGR is zero for GA, indicating that it is infeasible to satisfy the minimum rate constraint. Although our network can generate thousands of EPR pairs per second, satisfying the 10 EPRps minimum rate constraint requires a significant amount of network resources for distillation to meet the minimum fidelity threshold for each of the 10 EPR pairs. Therefore, using multiple paths can enhance the results. However, in a scenario where the set of user pairs is changing over time, and different paths need to be identified and installed for each user pair, the overhead of having multiple paths for each user pair could become a bottleneck. This scenario is similar to the problem of network updating in classical networks \cite{jin2014dynamic,reitblatt2012abstractions,kim2015kinetic}, which we consider as a potential area of future research.
%With one path per user pair, it is hard to guarantee the minimum rate constraint (

\subsection{Distillation strategies}
\begin{figure}[t]
\centering
\includegraphics[scale=0.5]{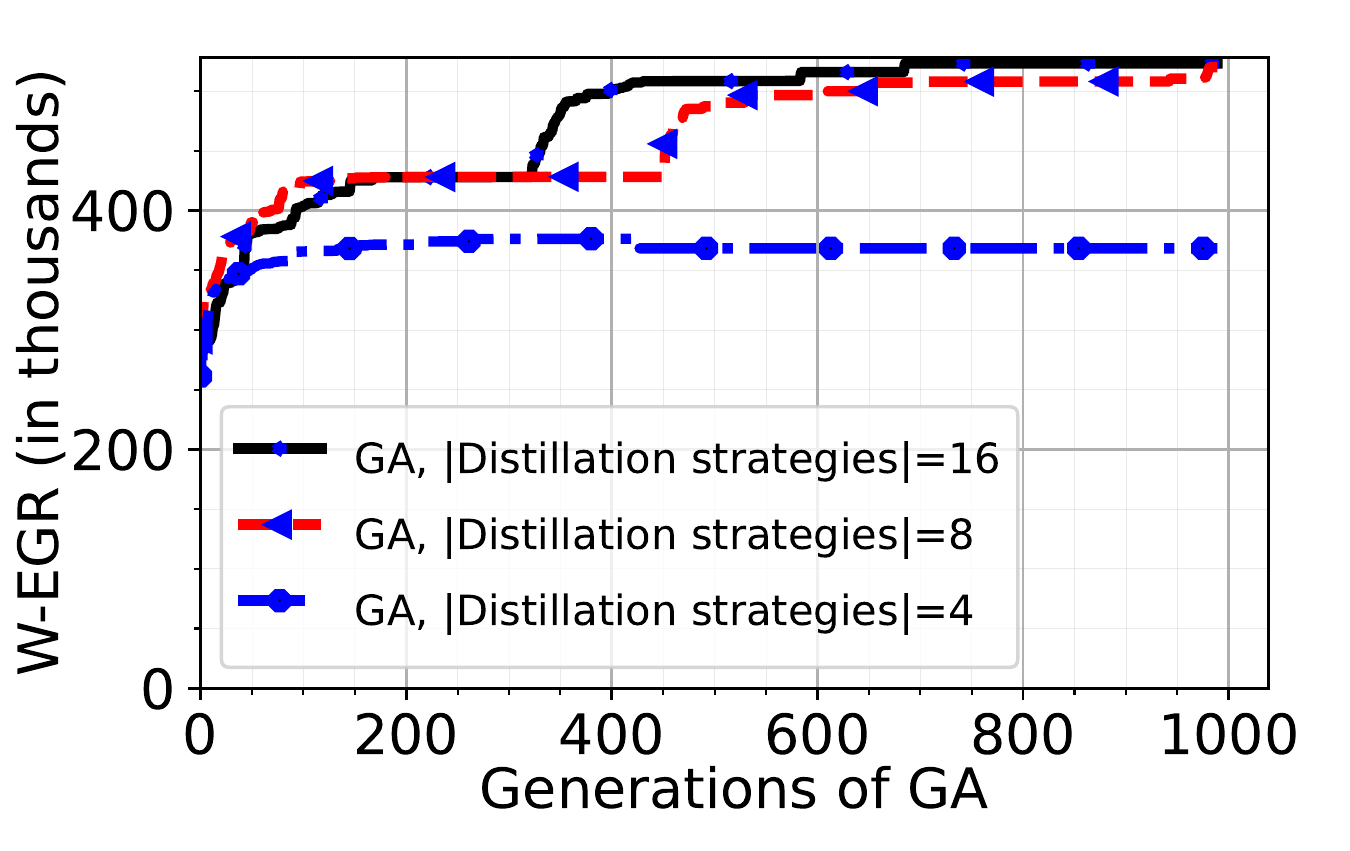}
\caption{Effect of the number of available distillation strategies on W-EGR.}
\vspace{-0.2in}
\label{fig:QVPN_GA_affect_of_search_space_distillation_strategies}
\end{figure}

This section evaluates the effectiveness of GA in selecting different distillation strategies for various paths. To test GA's ability to handle paths of diverse lengths, user pairs are randomly selected from anywhere in the network, regardless of distance. However, it is important to note that this approach may not result in a solution that satisfies the minimum rate constraint for user pairs. To address this issue, the minimum rate constraint is set to zero and the maximum rate constraint is set to infinity. Figure \ref{fig:QVPN_GA_affect_of_search_space_distillation_strategies} depicts the aggregate W-EGR across all organizations when GA has access to different numbers of distillation strategies for each path. As expected, GA can select paths with more distillation options more efficiently, resulting in increased aggregate throughput.

%In order to stress the GA and add in order to have different paths with more diverse lengths, we select our user pairs randomly from any place in the network no matter how far they are from each other. Obviously, in this way, there may not be any solution if we have a minimum rate constraint for user pairs. For that, we set the minimum rate constraint to zero and the maximum rate constraint value to infinity. Figure \ref{fig:QVPN_GA_affect_of_search_space_distillation_strategies} shows the aggregated weighted EGR across all organizations when the GA has access to a different number of distillation strategies for each path. As expected, when there are more options for distillation on a path, the GA can choose it as this will help to use the resources more efficiently and so the aggregate throughput will increase.
%With a higher number of available distillation strategies, GA is able to use the resources more efficiently.

\subsection{\textit{Gradient-based} scheme and GA vs. baseline algorithms}

We now evaluate our proposed \textit{Gradient-based} scheme’s speed and effectiveness in selecting paths in the network compared to GA and our baseline algorithms. For this experiment, we set up a single organization with $150$ user pairs and considered only one distillation strategy  (i.e., only end-level distillation) for each path. We did not set any minimum or maximum rate constraint for the user pairs, but we chose the fidelity threshold of user pairs uniformly in the range $[0.85,0.96]$. For this and the next experiment setup, we have used the default version of the SURFnet network (no repeater placement) with a multiplexing value of 4 for each link. We use the $k$-shortest path computed using three greedy-based heuristic schemes to build our candidate paths for GA and \textit{Gradient-based} scheme. $k$ in this experiment similar to other sections is $5$. The values of $\alpha$,  $\eta$, and $P_2$ are $0.1$, $1.0$ and $1.0$ respectively.

\begin{figure}
\centering
\includegraphics[scale=0.5]{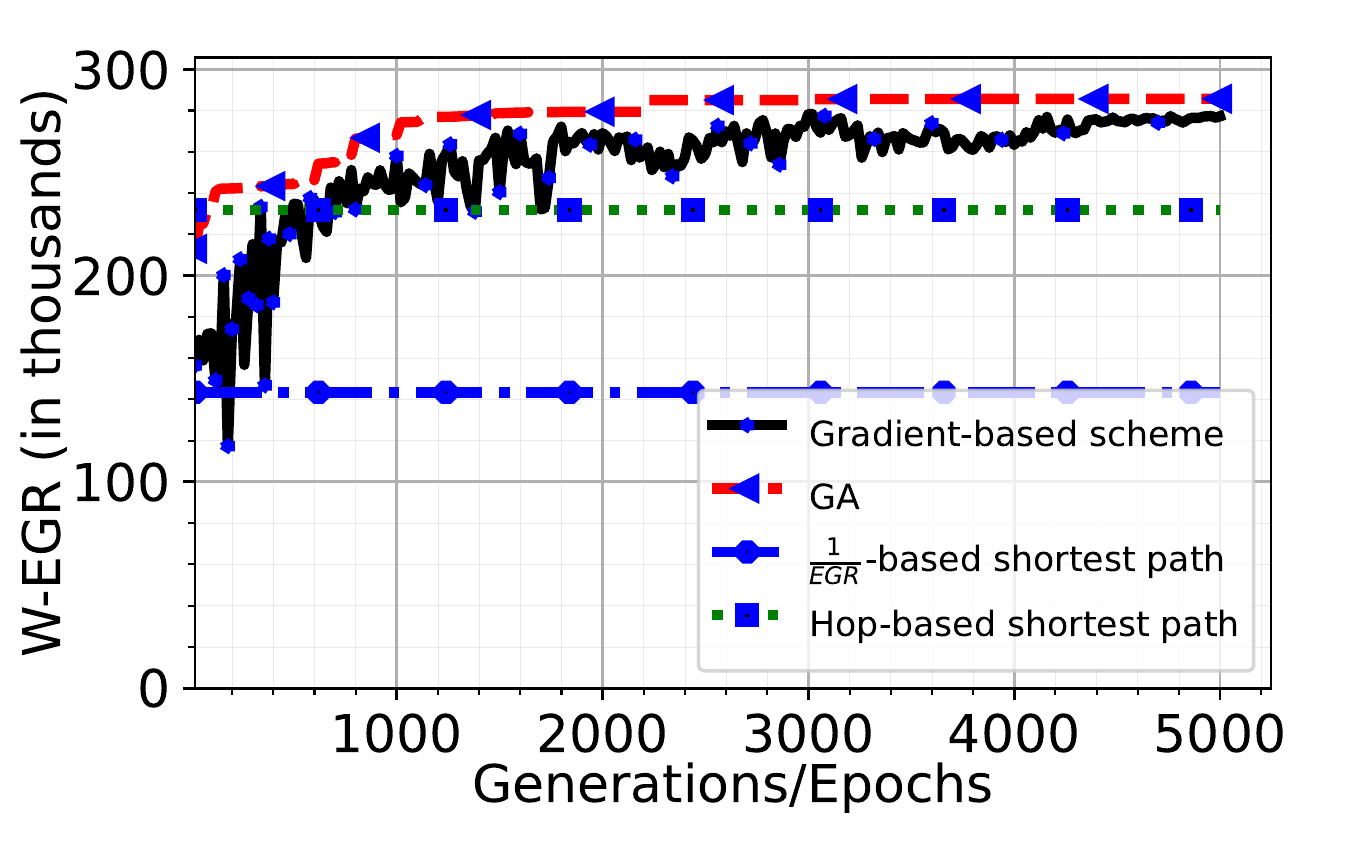}
\vspace{-0.1in}
\caption{\textit{Gradient-based} and GA schemes vs. baseline algorithms.}% The gradient-based scheme is much faster than the GA.}
\label{fig:QVPN_GA_GR_base_line}
\end{figure}

Figure \ref{fig:QVPN_GA_GR_base_line} shows the evolution of W-EGR in GA as a function of the number of generations and also the convergence of our \textit{Gradient-based} scheme as a function of training epochs. While these two terms are different, we have plotted them together for simplicity. We see that GA and the \textit{Gradient-based} scheme converge to almost the same solution and both over-perform the baseline algorithms. While the \textit{Gradient-based} scheme starts with a random solution (initializing the weights of the neural network randomly \ref{sec:appdx}) and so has a lower W-EGR at the beginning, it finds the best tuning for the neural network later and converges to what GA is able to. On the other hand, GA starts with a better solution as it has been initialized with our \textit{Hop-based} shortest path heuristic. The other observation is that there is some oscillation in the behavior of the \textit{Gradient-based} scheme over time and we do not have this behavior in GA. The reason is that GA has the ability to save the best solution from the previous generations and always replace the current best solution with a better solution in the newly generated population. On the other hand, the current implementation of our \textit{Gradient-based} scheme does not have this ability.

\subsection{Processing time of GA and \textit{Gradient-based} scheme}

As a follow-up, in this experiment, we evaluate the processing time of GA and compare it with the proposed \textit{Gradient-based} scheme. We have conducted this experiment on a Ubuntu $20.04.5$ LTS the Focal Fossa version with 6 CPUs and 8 GB RAM.   

\begin{figure}
\centering
\includegraphics[scale=0.5]{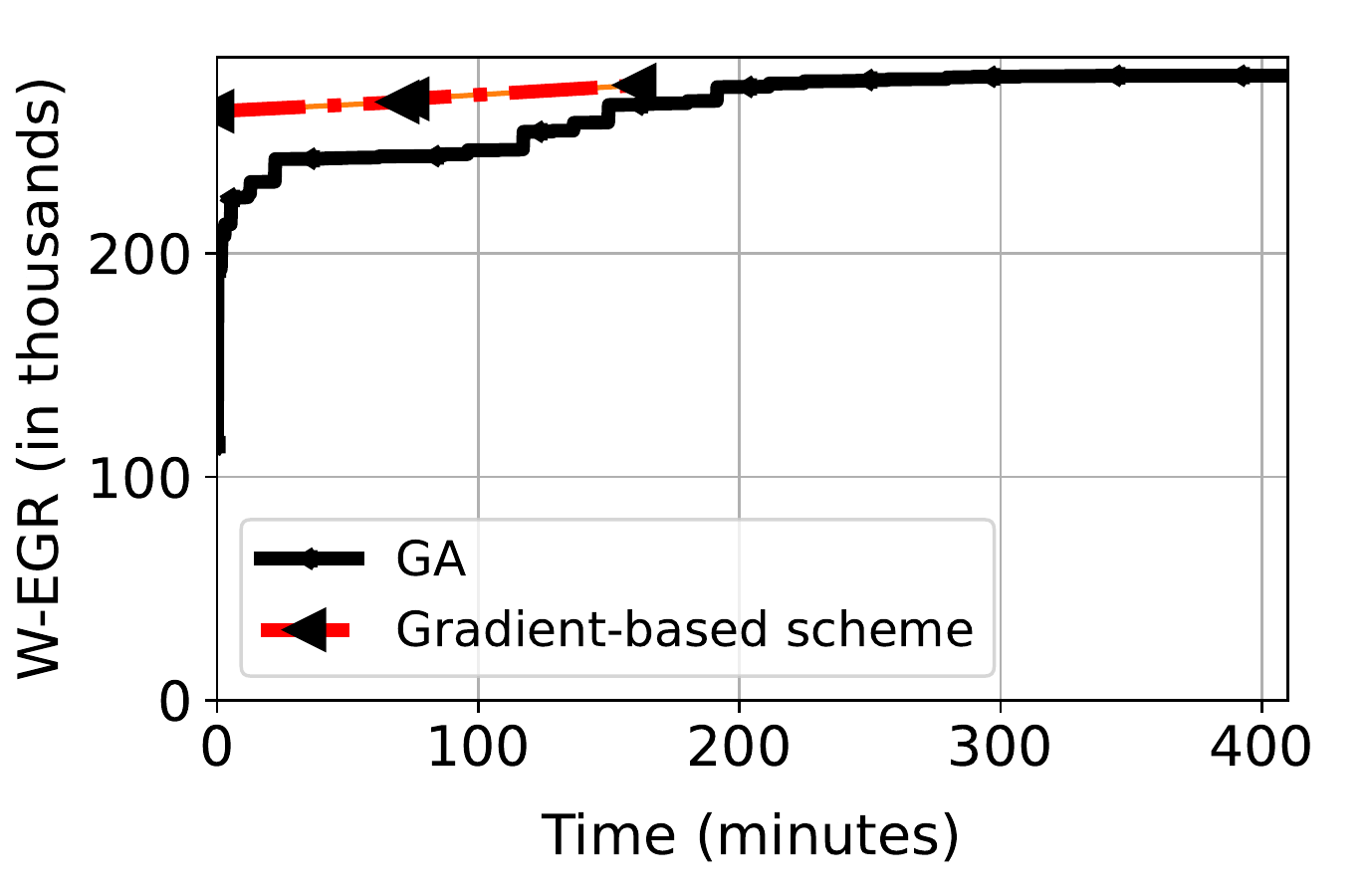}
\vspace{-0.1in}
\caption{Convergence of genetic and \textit{Gradient-based} algorithms.}% The gradient-based scheme is much faster than the GA.}
\label{fig:QVPN_GA_GR_as_time_in_minutes}
\end{figure}

Figure \ref{fig:QVPN_GA_GR_as_time_in_minutes} presents the aggregate W-EGR of the user pairs as a function of the processing time of GA and the \textit{Gradient-based} scheme in minutes. The population size for GA setup is $100$. The graph shows that the \textit{Gradient-based} scheme can find the same paths as the GA but with a significantly lower processing time. The conclusion from considering the results from figure \ref{fig:QVPN_GA_GR_base_line} and this figure is that while the \textit{Gradient-based} scheme needs more number of epochs to converge compared to the number of generations in GA, it has much less processing time for each epoch compared to GA processing time at each generation number. Basically, the processing for each epoch in the \textit{Gradient-based} scheme is to solve the optimization problem \ref{problem:flow_optimization} for all states in the batch and compute the gradients of the neural network and update its weights (more on this in section $\S$\ref{sec:appdx}). The size of the batch in the \textit{Gradient-based} scheme is negligible compared to the population size in GA. On the other hand, GA needs to evaluate all individuals (candidate solutions) in the current population and apply genetic algorithm operations to them to generate a new population. This process is time-consuming for big population sizes. One can reduce this processing time by reducing the population size in GA. But unfortunately, GA with smaller population sizes was not able to find good solutions for the problem we have in this paper.  

The downside of the \textit{Gradient-based} scheme is that its performance is sensitive to the problem and the output size. If we consider more distillation strategies for each path, the number of outputs in Figure \ref{fig:gb_for_path_selection} will be enormous. In this experiment, our  neural network has three layers, and the first layer is a convolutional layer with 128 filters. When we have a setup with multiple organizations and a minimum rate constraint, finding the optimal output sizes of the neural networks becomes crucial. We leave this as a potential area of future work.

\begin{figure}
\centering
\includegraphics[scale=0.5]{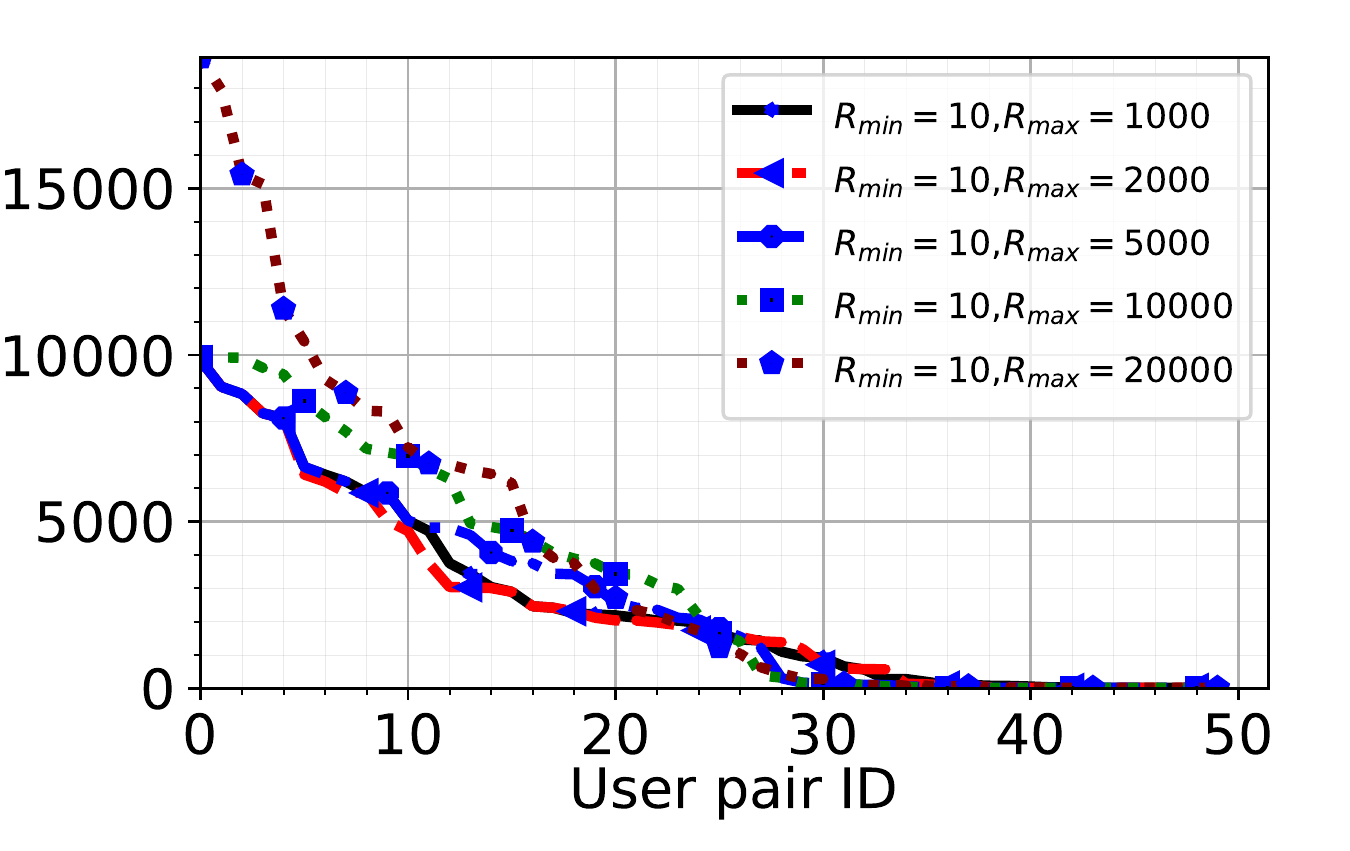}
\vspace{-0.1in}
\caption{True rates of each user pair in one organization with different maximum rate constraints.}% We can enforce a more fair allocation of resources to user pairs by setting a lower maximum rate constraint value, but this will reduce the aggregate throughput. }
\label{fig:fairness_among_flows}
\end{figure}

\subsection{Fairness among user pairs}

In this section, we examine the results of our qVPN in greater detail by analyzing the rates obtained by each user pair. While it may be expected that the optimization solver prioritizes user pairs with higher weights, we find that the length of the shortest paths used by each user pair and their fidelity threshold have a greater impact on the final rate that they receive from the network. To provide further insight into this, we present plots of the true EGR (unweighted) values received by each user pair under various minimum and maximum rate constraints.

In order to shape the traffic in the network, one approach is to vary the maximum rate constraints for each user pair. We set a minimum rate constraint of $10$ for all user pairs in each organization and a maximum rate constraint chosen uniformly at random from $[10, R_{max}]$.  We vary $R_{max}$ and use GA to optimize for the given configuration.

%We determine different minimum rate constraints for each user pair as follows. First, we find a set of paths for each user pair with a minimum rate constraint of 10 and a maximum rate constraint of 1000 for all user pairs in each organization. We use the shortest hop baseline to find a solution for this setup, and use the rate obtained for each user pair to set its minimum rate constraint. We then set the maximum rate constraint to different values and use the GA to optimize for that particular setup.

Figure \ref{fig:fairness_among_flows} depicts the true EGR obtained by each pair of users under different maximum rate constraints. The $x$-axis represents the user pairs sorted in descending order based on their true EGR. Several noteworthy observations can be made from Figure \ref{fig:fairness_among_flows}. As anticipated, decreasing the maximum rate constraint results in user pairs receiving more comparable rates from the network. However, the aggregate W-EGR decreases due to the lower maximum rate constraint. %We also observe that some user pairs, who have shorter path lengths, have the opportunity to receive a high rate.
%The first segment in the legends indicates the minimum rate constraint value and the second part indicates the maximum rate constraint value. The first segment on the first four schemes indicates we have gotten the value of the minimum rate constraint for each user pair from the solution of the original setup. 
%From figure \ref{fig:fairness_among_flows}, we can derive a couple of observations that are noteworthy. As expected, as we decrease the maximum rate constraint, user pairs get more similar rates from the network. However, the aggregated throughput is decreasing due to the lower maximum rate constraint. Some of the user pairs have the chance to get a high rate due to their shorter path length and maybe lower fidelity threshold. 

Table \ref{tab:correlation_table} displays the correlation coefficients that describe the association between a user pair's actual rate and the values of $\lambda^k_u$, $L_u^k$, $F^k_u$, and $\frac{w_k\lambda^k_u}{F^{k}_u*L_u^k}$. Here, $L_u^k$ is the length of the shortest path used by user pair $u$ belonging to organization $k$. We anticipate that user pairs with higher weights, lower fidelity thresholds, and shorter shortest path lengths will require fewer resources for distillation, making it easier to increase their EGR. The correlation coefficients reported in Table \ref{tab:correlation_table} for these metrics are consistent with this expectation. The correlation coefficient is a numerical value ranging between $-1.0$ and $1.0$ that reflects the degree of correlation between two entities. The value of the correlation coefficient for the true EGR and both $L_u^k$, $F^k_u$ are negative. This is in line with our expectation that user pairs with lower EPR fidelity requirements (lower $F^k_u$) would have higher rates, as fewer network resources would be allocated to distillation for those pairs. The correlation coefficient value between the true rate and $\frac{w_k\lambda^k_u}{F^{k}_u*L_u^k}$ is notably high and positive.

\begin{figure}
\centering
\includegraphics[scale=0.42]{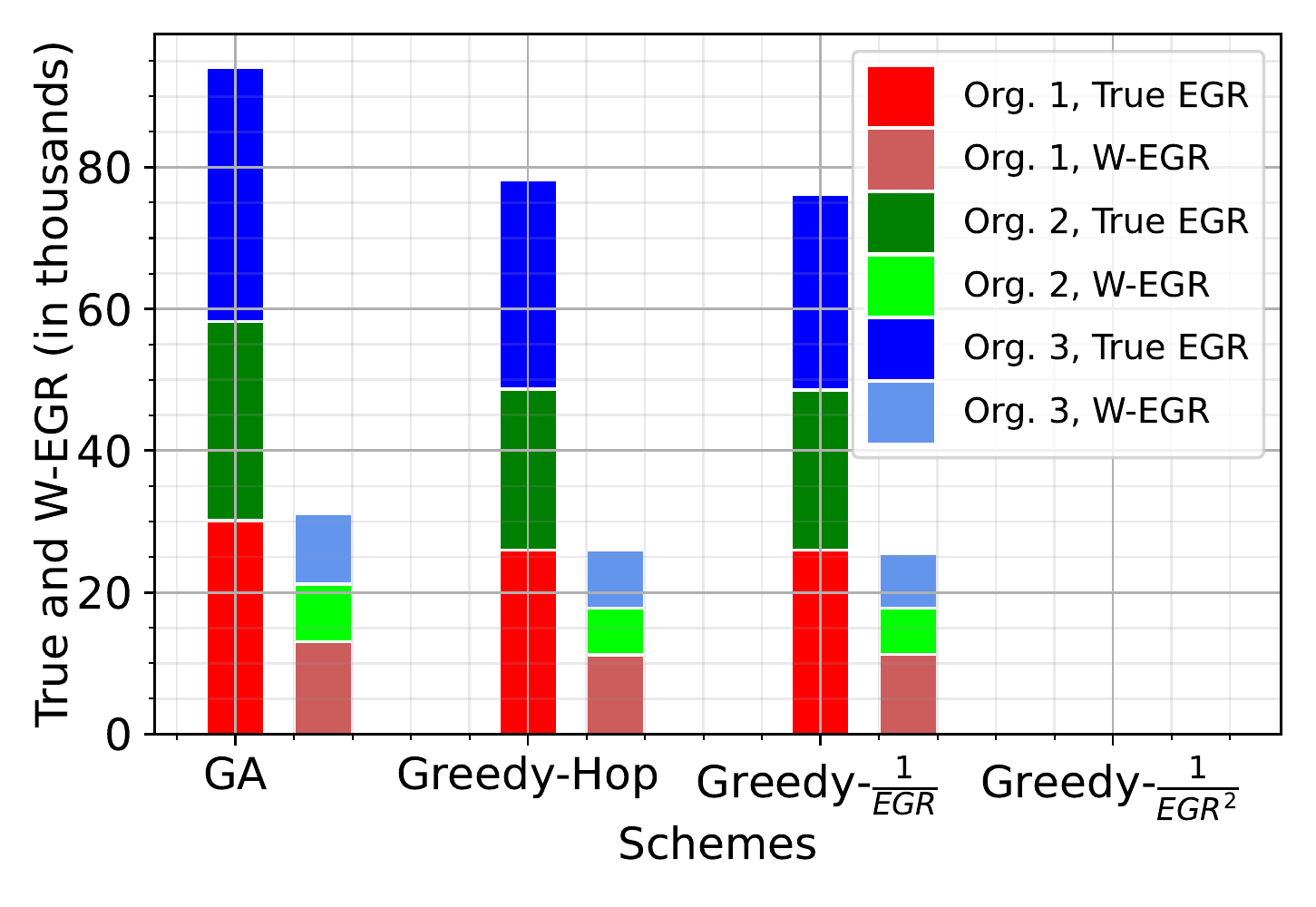}
\vspace{-0.1in}
\caption{True and weighted rates (W-EGR) delivered to each organization in different schemes.}% We can enforce a more fair allocation of resources to user pairs by setting a lower maximum rate constraint value, but this will reduce the aggregate throughput. }
\label{fig:delivered_rates_to_each_organization_stacked_barplotv1}
\end{figure}

We now examine the results of our experiments to explore how the resources are distributed among various organizations in the network following the application of the GA optimization method, that maximizes the aggregate W-EGR. Figure \ref{fig:delivered_rates_to_each_organization_stacked_barplotv1} shows the true and W-EGR for each organization in different schemes. We assigned a value of zero to the true and W-EGR for the Shortest EGR-Square-based scheme, since this scheme did not yield a viable solution using the suggested paths. The resources are allocated to the user pairs of each organization based on their weight. The higher the weight of the organization, the higher the true rate for that organization. It can be observed that GA surpasses all the baseline heuristics in terms of performance at the organization level.

%Since the weight for each organization and each user pair in each organization is selected uniformly in the range $[0.1,1.0]$ and $[0.3,0.7]$ respectively, each organization in every scheme is getting almost the same amount of E2E EPRs. The resources are allocated to the flows of each organization based on their weight. The higher the weight of the organization, the higher the true rate for that organization. 

%Figure \ref{fig:affect_of_shortest_path_on_true_rate} shows the relation between the true rate that each user pair in one of the organizations gets and the value of $\frac{Weight}{F^{th}*S.P.Lenght}$  where $Weight$ and $F^{th}$ are the weight and  the fidelity threshold of the user pair respectively and $S.P.Lenght$ is the length of the shortest path used by that user pair. We observe that the lower the value of ($\frac{Weight}{F^{th}*S.P.Lenght}$) for a user pair, the lower the true rate that it gets. This is understandable and can be interpreted as a user pair has a higher weight and a lower fidelity threshold and a smaller length for its shortest path, you need less amount of resources for distillation for that user pair and so it is more understandable to increase the rate of that user pair. %The other parameter that can affect the pattern is the length of the links that are used by each user pair and their capacity that we have not explored it here. 

\begin{table}[t]
\footnotesize
\begin{small}
    \centering
    \small
    \footnotesize
    \begin{tabularx}{\columnwidth}{X c c c c }
      \toprule
      {Parameter} &   {$\lambda^k_u$} & {$F^k_u$}&$L_u^k$ &$\frac{w_k\lambda^k_u}{F^{k}_u*L_u^k}$ \\
    \midrule
      {Correlation value} & {$0.186$}  &  {$-0.163$} &$-0.636$& {$0.576$} 
    \\
      \bottomrule
    \end{tabularx}
    \vspace{-0.02in}
    \caption{\label{tab:correlation_table} Correlation coefficient for user pairs received true rates and their characteristics (parameters).}
    \vspace{-0.1in}
    \end{small}
\end{table}
%Correlation coefficient for user pairs received true rates and their characteristics (parameters). There is a high correlation between the received true rate and $\frac{w_k\lambda^k_u}{F^{k}_u*L_u^k}$.

% \begin{figure}
% \centering
% \includegraphics[scale=0.4]{img/affect_of_Fth_W_S_P_Length_on_true_ratev2.jpg}
% \vspace{-0.1in}
% \caption{The relation between the true rate of each user pair and $\frac{w_k\lambda^k_u}{F^{k}_u*L_u^k}$.}
% \label{fig:affect_of_shortest_path_on_true_rate}
% \end{figure}

% \begin{figure}
% \centering
% \includegraphics[scale=0.2]{img/QVPN_GA_with_minimum_rate_constraintv2.pdf}
% \caption{Convergence of genetic algorithm with minimum rate constraint}
% \label{fig:Convergence_of_ga_with_minimum_rate_constraint}
% \end{figure}

% \subsection{Minimum rate constraint}
% Figure \ref{fig:Convergence_of_ga_with_minimum_rate_constraint} shows that the genetic algorithm is able to find a feasible solution and provide slightly higher aggregate weighted EGR compared to the heuristic schemes.
%\label{evaluation}

\section{Related Work}
\label{sec:related}
In this section, we overview the current state of research on resource management in quantum networks.
%\myparab{Routing and path selection in quantum networks}
While previous state-of-the-art studies \cite{chakraborty2019distributed,schoute2016shortcuts,pant2019routing,shi2020concurrent,Li2021,caleffi2017optimal} have mainly concentrated on developing routing protocols and path selection in quantum networks, there has been limited attention given to addressing resource management challenges related to entanglement distribution among multiple end-user pairs. Authors in \cite{da2021optimizing} introduce a method for optimizing the generation and distribution of entanglement through a combination of genetic algorithm and simulations of a linear quantum network. \cite{cicconetti2021request} addresses the problem of scheduling in the context of entanglement distribution. The authors have proposed a universal framework consisting of heuristic algorithms that prioritize minimizing application delay while also maximizing the entanglement rate and fidelity of E2E EPRs. The authors in \cite{Li2021} have proposed a centralized multi-path algorithm for quantum lattice networks with limited link capacities, while also ensuring a minimum E2E fidelity via only link-level purification. The proposed algorithm first identifies $k$ shortest paths for each user pair and then assigns link-level EPRs to paths based on concepts borrowed from classical network literature, such as proportional share and progressive filling. In \cite{victora2020purification}, the authors have developed entanglement distribution and purification protocols aimed at maximizing the distillable entanglement rate in a grid quantum network. For path selection, the authors propose three different strategies by assigning costs to links in the network and utilizing Dijkstra's algorithm to identify the shortest path for each end-user pair. In \cite{chakraborty2020entanglement}, authors consider a set up with multiple end user pairs, and optimize the distribution rate of EPR-pairs between them using an efficient linear programming formulation subject to minimum E2E fidelity constraint. However, the authors do not consider entanglement distillation and remove the paths from the candidate set which do not meet E2E fidelity requirement. None of the above mentioned works consider a multiorganizational setting with a share public quantum network infrastructure. Moreover, little focus has been given to access the fairness of the proposed resource allocation solution among different end-user pairs.

\section{Conclusion and Future Work}
\label{sec:concl}
In this paper, we introduced qVPNs and provided a thorough analysis and evaluation of their design. While VPNs are a well-established concept in the classical setting, we present the first comprehensive design and problem formulation of qVPNs. The main contributions of this work include formulating the resource allocation problem in qVPNs and demonstrating the ability to meet user specified QoS requirements using genetic algorithm and reinforcement learning based heuristics. Going forward, we plan to examine how memory decoherence affects the performance of qVPN. Additionally, we intend to explore situations where the set of user pairs is dynamic and changing over time.
% \begin{itemize}
%     \item Control plane scalability 
% \item Data plane scalability 
% \item Security plane scalability 
% \item Control protocol protection.

% \end{itemize}
\section{Acknowledgements}
 % \vspace{-0.14in}
This research was supported in part by the NSF grant CNS-1955744, NSF-ERC Center for Quantum Networks grant EEC-1941583, and MURI ARO Grant W911NF2110325. 
% \input{appendix}
%\newpage

{\small{
\bibliographystyle{plain}
\bibliography{sample-bibliography} 
}}
\section{Appendix}
\label{sec:appdx}
\subsection{Gradient-based reinforcement learning for path selection}
{\SetAlgoNoLine
\begin{algorithm}
  \hrulefill
  \vspace{-0.08in}
\caption{Gradient-based path selection algorithm}\label{alg:training_algorithm}

  \begin{algorithmic}[1]
  \STATE Randomly initialize policy network with weights $\theta$\\
  \STATE $n = \{\}$: Tracking the number of times we visit each state\\
  \STATE $v = \{\}$: Tracking the average reward from each state\\
  \STATE \For{each epoch}
  {
\STATE  \quad \quad  Sample $B$ repeated states (e.g., list of user\\
\quad \STATE \quad \quad pair IDs)
\quad  \quad \STATE \quad \quad \textbf{for $t=1$ to $|B|$ do}
    \quad  \quad  \quad   \STATE \quad \quad  \quad \quad  Sample  an action $a_{t,R}$ (based on policy\\    \quad \quad  \quad \quad $\pi$) for state $s_t$ 
  \quad \STATE \quad \quad  \quad \quad $n[s_t]++$ 
  \quad \STATE  \quad \quad \quad \quad Execute action $a_t$ and observe the \\ \quad \quad \quad \quad reward $r_t$ and transfer to the new state \\
  \quad \quad \quad \quad  $s_t=s_{t+1} \quad$(this state could be a new or \\
\quad \STATE \quad  \quad \quad \quad current list of user pair IDs.)
  \quad \STATE \quad \quad \quad \quad if $s_t \in v$:
  \quad \STATE \quad \quad \quad  \quad
    \quad $v[s_t] = v[s_t]+r_t$
    \quad \STATE \quad \quad \quad  \quad
  \quad $b(s_t)=\frac{v[s_t]}{n[s_t]}$
  \quad \STATE \quad \quad \quad \quad else:
  \quad \STATE \quad \quad \quad \quad \quad $v[s_t]=r_t,n[s_t]=1$\\
%   Append the state, reward and action to \\
%   \quad \quad \quad \quad the batch $B$\\
  \quad \quad \textbf{end}\\
  \quad \quad 
  \textbf{for} each state $s_t$ in batch $B$ \textbf{do}
  {
  \quad \STATE \quad \quad \quad \quad  $\Delta\theta= \Delta\theta + \alpha (\Delta\theta \log \pi (a_{t,R}|s_t;\theta)(r_t-$
  \STATE \quad \quad \quad \quad $b(s_t))+\beta\triangledown_{\theta} H(\pi(.|s_t;\theta)))$}
  \quad \STATE \quad \quad \quad \quad  Update the parameters: $\theta = \theta +\Delta \theta$\\
  \quad \quad \textbf{end}
  }
  \end{algorithmic}
\end{algorithm}
}
% \begin{figure}
% \centering
% \includegraphics[scale=0.4]{img/policy_network}
% \caption{Policy network architecture (described in $\S$\ref{sec:training_algorithm})}
% \label{fig:policy_nework}
% \end{figure}
% \vspace{-0.2in}

%\myparab{Learning algorithm:}
We describe our learning algorithm for path selection as follows.  A state in this algorithm indicates the list of user pairs. This algorithm can be used to learn a policy that outputs different probabilities for different paths based on the list of input user pairs. We use a policy gradient method \cite{sutton1999policy} to train the policy function. First, the algorithm randomly initializes all the weights $\theta$ of the policy network; (line 1). We update these parameters over $T$ epochs. For each input state $s_t$ (a list of user pairs) at epoch time $t$, the neural network selects action $a_{t,R}$ based on a policy $\pi$. Since $R$ different paths are sampled for each input state $s_t$, we define a solution $a_{t}^{R}=(a_{t}^{1} , a_{t}^{2} ,...,a_{t}^{R})$ as a combination of $R$ sampled actions using a stochastic policy $\pi(a_{t}^{R}|s_t)$ parameterized by $\theta$.  We execute the action and compute the reward (W-EGR). We use the state, action, and reward to update the weight of the network (lines 8-13). 

For each state $s_t$ in the batch $B$, we use the average reward of the state (computed at line 13) to update the weights of the neural network. In lines 16-17, we approximate a stochastic policy $\pi(a_{t} |s_t)$ parameterized by $\theta$ for selecting a solution $a_{t}$ for a given state $s_t$. The approximation is  as follows:

$$\pi (a_{t_{R}}|s_t;\theta) \approx \prod_{i=1}^{R} \pi(a^{i}_{t}|s_t;\theta).$$

In order to maximize the expected reward ($\mathbb{E}{r_t}$), we use gradient ascend using REINFORCE algorithm \cite{sutton1999policy,silver2014deterministic} with a baseline ($b(s_t)$). For each state $s_t$, we use an average reward as the baseline. The policy parameter $\theta$ is updated with learning rate $\alpha$ as following:

$$\theta = \theta +\alpha \sum_{t}^{} \nabla_\theta \log \pi(a_{t_{R}}|s_t; \theta)(r_t-b(s_t)).$$

In the above equation, for a given state $s_t$, we use $(r_t - b(s_t))$ to check how much better a specific solution is compared to the average
solution (our baseline solution). Note that we have used $\alpha$ in the paper as a parameter that controls the fidelity and the rate of each link also.  When $r_t > b(s_t)$, the probability of the solution $a_{t_{R}}$ is increased by updating policy parameters $\theta$ in the direction  $\nabla_\theta \log \pi(a_{t_k}|s_t; \theta)$. The update step size is $\alpha(r_t - b(s_t))$.
Otherwise, the solution probability is decreased. In this way, we reinforce actions that lead to better
rewards. We use entropy factor $\beta=0.1$ (strength of the entropy regularization term) in the training phase in order to ensure that the agent explores the action space adequately to discover good policies.

% \newpage
% \newpage
%\input{unprocessed_content}

\end{document}